\documentstyle[multicol,prb,aps,psfig]{revtex}

\input epsf

\def\prb{Phys. Rev. B}
\def\prl{Phys. Rev. Lett.}

\def\be{\begin{equation}}
\def\ee{\end{equation}}
\def\ba{\begin{eqnarray}}
\def\ea{\end{eqnarray}}

\def\LSCO{La$_{2-x}$Sr$_x$CuO$_4$}

\def\YBCO{YBa$_2$Cu$_3$O$_{7-\delta}$}

\def\BSCO{Bi$_2$Sr$_2$CaCu$_2$O$_8$}
\def\C60{A$_x$C$_{60}$}

\def\hts{high temperature superconductors}

\def\S{  {\cal S} }

\begin{document}


\title
{Dimensional Crossover in Quasi One-Dimensional and 
High $T_c$ Superconductors}

\author{E.~W.~Carlson,  D.~Orgad}
\address
{Dept. of Physics,
U.C.L.A.,
Los Angeles, CA  90095}
\author{S.~A.~Kivelson}
\address
{Dept. of Physics and Department of Applied
Physics, Stanford University, Stanford, CA 94305-4045
and Dept. of Physics
U.C.L.A.,
Los Angeles, CA  90095}
\author{V.~J.~Emery}
\address{
Dept. of Physics,
Brookhaven National Laboratory,
Upton, NY  11973-5000}
\date{\today}
\maketitle 

\begin{abstract}

The one-dimensional electron gas exhibits spin-charge separation and 
power-law spectral responses to many experimentally relevant probes.  
Ordering in a quasi one-dimensional system is
necessarily associated with a dimensional crossover, 
at which sharp quasiparticle
peaks, with small spectral weight, emerge from the incoherent background.  
Using methods of Abelian bosonization, we
derive asymptotically correct expressions for the spectral changes 
induced by this crossover. Comparison is made with experiments on the 
high temperature superconductors, which are electronically quasi 
one-dimensional on a local scale.

\end{abstract}

\begin{multicols}{2}


In this paper, we consider the spectral signatures of dimensional 
crossover in the continuum theory of a quasi one-dimensional 
superconductor.  This problem is of interest in its own right and for 
application to materials which are structurally quasi one-dimensional, 
such as the Bechgaard salts (organic superconductors).  We believe that it 
is also interesting as a contribution to the theory of the high 
temperature superconductors. Although structurally these materials 
are quasi two-dimensional, there is both   
theoretical and experimental evidence\cite{pnas} of a substantial range of 
temperatures in which ``stripe'' correlations make the electronic structure 
locally quasi one-dimensional,
a phenomenon we have labeled 
``dynamical dimension reduction.'' Similarly, the ET versions of organic
superconductors are two-dimensional doped antiferromagnets, which we expect
to show similar behavior. More generally,
the high temperature superconducting state emerges from a non-Fermi 
liquid normal state, often with a normal-state pseudogap. The 
quasi one-dimensional superconductor is the only solvable case in 
which such an evolution can be traced, theoretically.  

A quasi one-dimensional system can be thought of as an array of 
``chains'', in which the electron dynamics within a chain is characterized 
by energy scales large compared to the 
electronic couplings between chains.  Since a one-dimensional system 
cannot undergo a finite temperature phase transition, any ordering 
transition with a finite critical temperature $T_{c}$ is necessarily 
associated with a dimensional crossover. The electronic properties at
temperatures (or energies) large compared to $T_{c}$ can be 
understood by ignoring the interchain coupling, while at lower 
temperatures or energies, the behavior is that of a three 
dimensional  system.  

In the one dimensional electron gas\cite{review} (1DEG), 
as a consequence of spin-charge separation,
the elementary excitations are collective modes with unusual quantum
numbers and topological properties:  The charge excitations are
best understood as 
sound-like density-wave phasons (or, in dual representation, superconducting 
quasi-Goldstone modes) when the system is gapless, and as charge solitons with 
charge $\pm 1$ and spin 0 when a charge gap is induced.  
(The precise meaning of the soliton ``charge" is a quantized unit of 
chirality; 
See Eq. (\ref{eq:Nc}) and the subsequent
discussion.) Similarly, the spin  excitations of a spin-gapped system 
are spin solitons with charge 0 and spin $1/2$.  
When the elementary excitations do not have the quantum numbers of the 
experimentally accessible excited states,  
spectral functions do not exhibit sharp peaks corresponding 
to a well defined mode with a definite dispersion relation,
$\omega=\epsilon(k)$.  The single hole spectral function, $G^{<}(k,\omega)$, 
which is measured in angle resolved photoemission spectroscopy (ARPES), 
involves excited states with charge $e$ and spin 1/2, 
which thus consist of at least one charge soliton
and one spin soliton.  The dynamic spin structure factor $\S(k,\omega)$, 
measured by neutron scattering, involves excited states with spin 1, 
which thus consist of two spin solitons.
(We will see in Section \ref{sec:lowt}C that, in fact, the 
relevant excited states
contain two spin solitons and at least two charge antisolitons.)

Below $T_{c}$, where the system is three dimensional, we will show
that the solitonic excitations of the 1DEG are confined in multiplets
with quantum numbers that are simply related to those of the electron.  For
the case of three dimensional charge-density wave ordering, this has
been known for some time.  For the case of the superconductor, it is
related to the fact, noted recently by Salkola and Schrieffer,\cite{salkola}
that either a spin soliton or a charge soliton induces a $\pi$-kink 
in the superconducting 
correlations.  As a consequence of confinement, there is a finite probability
of creating a final state consisting of a single bound spin and charge 
soliton pair 
in an ARPES experiment. This will show up as a coherent (delta-function) 
piece in the zero temperature $G^<(k,\omega)$. 

In this paper, we show that the coherent piece
of the single particle spectral functions has a weight which vanishes in the 
neighborhood of $T_c$ in proportion to a positive power of the interchain 
Josephson energy.   
It is this fact, that the {\it spectral weight} of the coherent piece
is strongly temperature dependent below $T_c$, rather
than either the energy or the lifetime of the normal mode, 
which is a new feature that emerges from our analysis.  
It is highly reminiscent of behavior observed in 
ARPES\cite{fedorov,shen} and inelastic neutron scattering\cite{neutron} 
measurements on the high temperature superconductors.
We have also identified a new resonant feature in the
spin spectrum of a quasi one-dimensional superconductor that emerges 
at temperatures well below $T_c$.

If the 1DEG remains gapless down to  $T_c$,
the superconducting transition is BCS-like, in the sense that both pairing and
phase coherence occur at the same time. In this case both are induced by
the interchain Josephson tunnelling.
We will mainly be concerned with the case in which
a sort of ``pairing'', {\it i.e.} the opening of a spin gap
$\Delta_s > 0$, occurs in the 1d regime well above $T_c$. In this case 
$T_c$ is primarily associated with phase ordering, and its scale is set by the
superfluid density,\cite{uemura,phase} 
rather than by the zero temperature single particle gap scale, $\Delta_0/2$. In such 
circumstances the superconducting state, even at very low temperatures,  
maintains a memory of the
separation of charge and spin which is a feature of the (1d) normal state.  
The unique ``coherence length'' of a BCS superconductor is replaced by two
distinct correlation lengths\cite{finkelstein}: a spin length, 
$\xi_s=v_s/\Delta_s$, where $v_s$ is the
spin velocity, and a charge length, 
$\xi_c=v_c/\Delta_c$, where $\Delta_c\sim$ 2$T_c$.

The remainder of the paper is divided into two self-contained parts;  
in Sections \ref{sec:bos}-\ref{sec:lowt} we derive asymptotically 
exact results for the spectral properties of a quasi one-dimensional 
superconductor in the limit of weak interchain coupling. 
In Section \ref{sec:summary}, we summarize the principal results and
discuss their  application to experiment, especially in the high temperature
superconductors. 
The reader who is interested only in results, not their derivation, 
can skip the intervening sections.

The model we study is defined by the Hamiltonian
\be
H=\sum_j\int dx  {\cal H}_j + H_J \; ,
\label{eq:H}
\ee
where the sum runs over chains, $H_j$ is the Hamiltonian of the 
1DEG on chain $j$,
and $H_J$ is the Josephson coupling between
chains.  In Secs. \ref{sec:bos}-\ref{sec:intt}, 
we consider the single chain problem ($H_J=0$).
The problem is formulated using Abelian bosonization in Sec. \ref{sec:bos}.
Next, we discuss the spectral functions for the 
1DEG without (Section \ref{sec:hight}) and with (Section \ref{sec:intt}) 
a spin gap - explicit expressions for 
various quantities in the presence of a spin gap are reported here for 
the first time of which we are aware.  In Section \ref{sec:lowt}, 
we extend these results to the case in which the most relevant 
interchain coupling is the Josephson tunnelling.
An adiabatic approximation, which is exact in the limit where 
$\Delta_{s}\gg\Delta_{c}$, replaces the spin-charge separation of 
the purely 1d problem as the central feature of the spectrum
- this section contains our principal new results. 
Applications to {\hts} are described in Sec. \ref{sec:summary}. 
Various appendices expand upon the derivations in Section \ref{sec:lowt}.

\section{Abelian Bosonization and the Spectral Functions}
\label{sec:bos}

We begin by considering the properties of a single chain in the absence of
any interchain coupling;  we treat this problem using Abelian bosonization, 
which is based on the fact that the properties of an  
interacting 1DEG at low energies and
long wavelength are asymptotically equal to those of a set of two independent
bosonic fields, one representing the charge and the other the spin degrees of
freedom in the system.  The widely discussed separation of charge and spin 
\cite{LE,ELP} in this problem is formally the statement that the Hamiltonian 
density ${\cal H}_j$ can be expressed as
\be
{\cal H}={\cal H}_c+{\cal H}_s \; ,
\ee
where the chain index is implicit, and the charge and spin pieces of the 
Hamiltonian are each of the sine Gordon variety, 
\be
{\cal H}_{\alpha} = \frac {v_{\alpha}} 2\left [ 
K_{\alpha}(\partial_{x}\theta_{\alpha})^{2}
+\frac {(\partial_{x}\phi_{\alpha})^{2}} {K_{\alpha}}
\right ] + V_{\alpha} \cos(\sqrt{8\pi} \phi_{\alpha}) \; ,
\nonumber
\ee
where $\alpha = c,s$ for the charge and spin fields, respectively,
$\theta_{\alpha}$ is the dual field to $\phi_{\alpha}$, or equivalently
$\partial_{x}\theta_{\alpha}$ is the momentum conjugate to 
$\phi_{\alpha}$.  We consider a sufficiently incommensurate 1DEG and 
therefor set $V_{c}=0$ since it
arises from Umklapp scattering. Of course, if the Umklapp scattering
is crucial to explain doped insulator behavior, its role cannot be
neglected. Where there is no spin gap, or at temperatures
large compared to $\Delta_{s}$, we can likewise set $V_s=0$. 

When $V_s$ is relevant, (perturbatively, this means $K_s <1$) the
spin gap is dynamically generated, {\it i.e.} it depends both on $V_{s}$ 
and the ultraviolet cutoff in the  problem, $\Lambda$, according to the 
scaling relation 
$\Delta_{s}\sim v_s\Lambda[V_s/v_s\Lambda^2]^{1/(2-2K_s)}$. 
At the gapless fixed point, spin-rotational
invariance requires $K_s = 1$, at which point $V_s$ is perturbatively 
marginal.  It
is marginally irrelevant for repulsive interactions 
($K_s>1$) and
marginally relevant for attractive interactions 
($K_s<1$).  
Thus, the long distance spin physics is described by ${\cal H}_{s}$ with 
$V_s=0$ and $K_s=1$ for a gapless spin-rotationally invariant phase.  Where 
there is a spin gap in a spin rotationally invariant system, it is 
exponentially small for weak interactions,
$\Delta_s \sim \sqrt{V_sv_s}\exp[-v_s\Lambda^2 /2\pi V_s]$.

In order to compute correlation functions, we use the Mandelstam
representation\cite{sm} of the fermion field operators 
\be
\psi_{\lambda,\sigma}(x)={\cal N_{\sigma}}  \exp\left[ i\lambda k_{F}x 
-i\Phi_{\lambda,\sigma}(x)\right ] \; ,
\label{eq:bose}
\ee
where
${\cal N_{\sigma}}$ contains both a
normalization factor (which depends on the ultraviolet cutoff) and a 
``Klein'' factor (which can be implemented in many ways) so that 
${\cal N_{\sigma}}$ anticommutes with ${\cal N_{\sigma^{\prime}}}$
for $\sigma \ne {\sigma^{\prime}}$ and commutes with it for $\sigma =
{\sigma^{\prime}}$. In addition
\be
\Phi_{\lambda,\sigma}=
\sqrt{\pi/2}\left[(\theta_{c}-
\lambda  \phi_{c}) + \sigma( \theta_{s}
-\lambda  \phi_{s})\right] \; ,
\label{eq:Phi}
\ee
where $\lambda = -1$ for left moving electrons, $\lambda = +1$ for
right moving electrons, and 
$\sigma=\pm 1$ refers to spin polarization.   
From Eq. (\ref{eq:bose}), it is a straightforward 
(and standard\cite{review}) exercise to
obtain the boson representations of all interesting electron bilinear 
and quartic operators.
Physically, $\phi_c$ and $\phi_s$ are,
respectively, the phases of the
$2k_F$ CDW and SDW fluctuations, and $\theta_c$ is the 
superconducting phase. 
The long-wavelength components of the charge
($\rho$) and spin ($S_z$) densities are given by
\ba
\rho(x)=&&
\sum_{\lambda,\sigma}\psi^{\dagger}_{\lambda,\sigma}\psi_{\lambda,\sigma}
-\frac{2k_F}{\pi}=\sqrt{\frac{2}{\pi}}\partial_x \phi_c \; ,
\nonumber \\ 
S_z(x)=&&
\frac{1}{2}\sum_{\lambda,\sigma}\sigma\psi^{\dagger}_{\lambda,\sigma}
\psi_{\lambda,\sigma}=\sqrt{\frac{1}{2\pi}}\partial_x \phi_s \; .\nonumber
\ea

When analyzing results for this model, it is always important
to remember that the parameters which enter the field theory are renormalized,
and are related to the microscopic interactions in a 
very complicated
manner.  For instance, although for a single-component 1DEG with repulsive
interactions $V_s$ is always irrelevant, for multicomponent 1DEG's, and for the
``1DEG in an active environment'', it is common to find a dynamically generated
spin gap, even when the microscopic interactions are uniformly
repulsive.\cite{spingap,spingap1,spingap2,spingap3}  

The bosonized expressions for all electron operators are readily 
extended to an array of chains by adding a chain index to the Bose 
fields and to the Klein factors;  the Klein factors on different 
chains must now anticommute with each other.  Where single particle 
interchain hopping is relevant, the Klein factors appear explicitly 
in the bosonized Hamiltonian. Where only pair hopping and 
collective interactions between neighboring chains need be 
included in the low energy physics, the Klein factors cancel in $H$.

While it is generally simpler to derive results concerning the spectrum, it is
important for comparison with experiment to compute actual correlation 
functions. Specifically, we will consider the transverse spin dynamic 
structure factor
\ba
\label{sstract}
\nonumber
\tilde{\cal S}(x,t;T)&\equiv&\langle{{\rm S}^x_{2k_F}}^
{\dagger}(x,t){\rm S}^x_{2k_F}(0,0)\rangle \\
&+&
\langle{{\rm S}^y_{2k_F}}^
{\dagger}(x,t){\rm S}^y_{2k_F}(0,0)\rangle \; , 
\ea
where
\be
{\bf S}_{2k_F}=\frac{1}{2}\sum_{\sigma,\sigma'}\psi_{1,\sigma}^{\dagger}
\mbox{\boldmath $\tau$}_{\sigma\sigma'}\psi_{-1,\sigma'} \; ,
\ee
and the $\mbox{\boldmath $\tau$}$ are Pauli matrices.
We will also consider the one-hole Green function,
\ba
\tilde G^{<}(x,t) \equiv && \langle
\psi_{-1,\uparrow}^{\dagger}(x,t)\psi_{-1,\uparrow}(0,0)\rangle \; ,
\ea
the singlet-pair correlator,
\be
\tilde \chi(x,t) \equiv \langle
\psi_{1,\uparrow}^{\dagger}(x,t)\psi_{-1,\downarrow}^{\dagger}(x,t)
\psi_{-1,\downarrow}(0,0)\psi_{1,\uparrow}(0,0)\rangle \; ,
\ee
and the various spectral functions, $\S$, $G^<$, and $\chi$, obtained by 
Fourier transforming these correlators.  As a consequence of separation 
of charge and spin, $\tilde \S$, $\tilde G^<$, and $\tilde \chi$ are 
expressible as a product of spin and charge contributions, and consequently 
$\S$, $G^<$, and $\chi$ are convolutions.  For instance,
\be
G^{<}(k,\omega)=\int {dq\over 2\pi}{d\nu\over2\pi} 
G_{s}(k-q,\omega-\nu)G_{c}(q,\nu) \; .
\ee

\section{ High temperature: Luttinger liquid behavior}
\label{sec:hight}

At temperatures large compared to $T_c$ and the spin gap, $\Delta_s$, 
(or at all
temperatures in systems in which
$T_c=\Delta_s=0$), the 1DEG exhibits ``Luttinger liquid''
behavior.  Because the Luttinger liquid is a quantum critical system, the
response functions have a scaling form.
Specifically, this implies\cite{review} that
\be
G^{<}(k,\omega;T)=T^{2\gamma_c+2\gamma_s-1}G^<(k/ T,\omega/T;1)\; , 
\ee
where we define for $\alpha=c$ or $s$
\be
\label{gammadef}
\gamma_{\alpha}=\frac{1}{8}(K_{\alpha}+K_{\alpha}^{-1}-2) \; ,
\ee
and that so long as $K_s\ge 1$,
\be
\S(k,\omega;T) = T^{(K_s^{-1}+K_c-2)}\S(k/T,\omega/T;1) \; .
\ee
Note that here, and henceforth, we will measure $k$ relative to $k_F$ and
$2k_F$, respectively, when computing the scaling functions $G^<$ and $\S$.  
If the system is spin-rotationally invariant, $K_s=1$ in the above
expressions.  

The form of these scaling functions can be computed analytically in
many cases;  this has recently been accomplished in Ref. \ref{ref:dror}. 
They may or may not have a peak at energies small 
compared to the bandwidth, depending on certain exponent 
inequalities.  Where there is a peak, it occurs at positive 
energies
$\omega=\pm v_{\alpha}k + ({\rm const.}) T$, but the peak width, however 
defined, does not narrow in proportion to $T$ at low temperatures;  
such a peak does {\it not} correspond to a quasiparticle.

\section{ Intermediate temperature:  The Luther-Emery liquid}  
\label{sec:intt}

When $V_s$ is
relevant, the spin sine Gordon theory scales to a strong-coupling fixed point, 
and the excitations are massive solitons, in which $\phi_s$ 
changes by $\pm\sqrt{\pi/2}$ ({\it i.e.} $S_z=\pm 1/2$).  This problem is most 
simply treated in terms of spin fermion fields,
\be
\Psi_{s,\lambda}^{\dagger}= {\cal 
N}_{s}\exp[i\sqrt{\pi/2}(\theta_{s}-2\lambda \phi_{s})].
\ee 
The refermionized form of the Hamiltonian is then
\ba
{\cal H}_{s}= && i\tilde v_{s}[\Psi_{s,-1}^{\dagger}\partial_{x}\Psi_{s,-1}-
\Psi_{s,1}^{\dagger}\partial_{x}\Psi_{s,1}] \nonumber \\
&& + \tilde \Delta_{s}
[\Psi_{s,1}^{\dagger}\Psi_{s,-1}+{\rm H.c.}] \nonumber \\ && +
g_s\Psi_{s,1}^{\dagger}\Psi_{s,-1}^{\dagger}\Psi_{s,-1}\Psi_{s,1} \; ,
\ea
where 
\begin{eqnarray}
\nonumber
\tilde v_s&=&v_s\left(\frac{1}{4K_s}+K_s\right) \; , \\
\nonumber
\tilde \Delta_s&=&\frac{\pi V_s}{\Lambda} \;  , \\
g_s&=&2\pi v_s\left(\frac{1}{4K_s}-K_s\right) \; .
\end{eqnarray}

For $K_s=1/2$, which is known as the Luther-Emery point,\cite{LE} 
the refermionized model
is non-interacting and massive,  
with a gap $\Delta_s = \tilde \Delta_s$.  Assuming there is a 
single massive phase of the sine Gordon
theory, the Luther-Emery model\cite{LE} will exhibit the same asymptotic
behavior as any other model in this phase.  
Formally, the Luther-Emery point can be thought of as a
strong-coupling fixed point Hamiltonian, and $g_s$, which 
vanishes at the fixed point, is the amplitude
of a leading irrelevant operator.\cite{ideology}  
We will henceforth compute correlation functions at the Luther-Emery
point, and then comment on the effects of deviations from this point.

Now, in computing the various spectral properties of the system, 
we can distinguish two regimes of temperature:  
at temperatures large compared to $\Delta_s$, the spin gap is
negligible, and the results for the Luttinger liquid apply.  
If the temperature is small compared to the spin gap, then we can compute 
the spin contributions to the various
correlation functions in the zero-temperature limit, 
and only make exponentially small
errors of order $\exp(-\Delta_s/T)$.  
The spin piece of the transverse spin 
response function can be expressed in terms of the spin fermion fields
\be
\tilde \S_{s}(x,t) = \langle\Psi_{s,1}^{\dagger}(x,t)\Psi_{s,-1}^
{\dagger}(x,t)\Psi_{s,-1}(0,0)\Psi_{s,1}(0,0)\rangle \; .
\ee
Since the theory reduces, at the Luther-Emery point, to a theory of free 
massive fermions, the corresponding spectral function can be readily computed 
with the result, for $T=0$,  
\be
{\cal S}_s(k,\omega)=\frac{\omega^2-4E_s^2(k/2)}{4v_s^2|q_1 E_s(q_2)
-q_2 E_s(q_1)|}\Theta[\omega-2E_s(k/2)] \; ,
\label{eq:Sspngap}
\ee
where the spin soliton spectrum is
\be
\label{es1def}
E_s(k)=\sqrt{v_s^2k^2+\Delta_s^2} \; ,
\ee
and $q_{1,2}$ are the solutions of the quadratic equation  
$\omega+E_s(q)+E_s(k+q)=0$. Explicitly
\be
\label{q1q2s}
q_{1,2}=\frac{k}{2}\pm\frac{\omega}{2v_s}\sqrt{1+\frac{4\Delta_s^2}
{v_s^2 k^2-\omega^2}} \; .
\ee

The spin piece of the one hole Green function is more complicated, 
since it involves nonlocal operators
in the refermionized form:
\be
\tilde G_{s}(x,t)=\langle U_s^{\dagger}(x,t)\Psi_{s,-1}^{\dagger}(x,t)
\Psi_{s,-1}(0,0)U_s(0,0)\rangle \; ,
\ee
where the vertex operator
$U_{s}(x)=e^{i\sqrt{\pi/2}\phi_s(x)}$ with 
$\phi_s(x) =\sqrt{\pi/2}\sum_{\lambda}\int^xdy
\Psi_{s,\lambda}^{\dagger}\Psi_{s,\lambda}$.  
From kinematics, it follows that 
this Green function consists of a coherent one spin soliton
piece and an incoherent multisoliton piece:
\ba
G_{s}(k,\omega)=  Z_s(k) \delta[\omega-E_{s}(k)]
 +G_{s}^{(multi)}(k,\omega) \; ,
\label{eq:Gs}
\ea
where the multisoliton piece is proportional to $\Theta[\omega-3E_{s}(k/3)]$.
(Deviations from the Luther-Emery point  
in the case $g_s>0$ will result in the formation of a
spin soliton-antisoliton bound state, a ``breather'', 
which can shift the threshold energy for
multisoliton excitations somewhat.) 

At the Luther-Emery point it is possible to obtain closed form 
expressions\cite{leclair} for the matrix elements of the
vertex operator between the vacuum and various multisoliton 
states and from that to 
compute $Z_s$ explicitly.  
We will report this calculation in a forthcoming paper, Ref.
\ref{ref:dror}.  Here, we use a simple scaling argument, 
which can be generalized to the
case of nonzero interchain coupling, 
to derive the principal features of this result, especially
the dependence of
$Z_s$ on
$\Delta_s$. In the absence of a spin gap, and at $T=0$, $G_s$ can be readily  
evaluated to give the scaling form
\ba
\nonumber
G_s=\frac{\pi\,(v_s\Lambda)^{\frac{1}{2}-2\gamma_s}}{\Gamma(\gamma_s)\Gamma
(\gamma_s+\frac{1}{2})}
(\omega+v_s k)^{\gamma_s-1}&&(\omega-v_s k)^{\gamma_s-\frac{1}{2}} \\
&&\!\!\!\!\!\!\times\Theta(\omega-v_s|k|) \; . 
\label{eq:gslutt}
\ea
Because the sine-Gordon field theory is
asymptotically free, the high energy spectrum, and hence the dependence 
of $G_s$ on $\Lambda$, is unaffected by the opening of a spin gap.  
With this observation, it is simply a matter of dimensional analysis to 
see that
\be
Z_s(k)=(\Lambda\xi_s)^{\frac{1}{2}-2\gamma_s} f_s(k\xi_s) \; , 
\label{eq:Zs}
\ee
where $\xi_s=v_s/\Delta_s$ is the spin correlation length. $f_s$ is a scaling 
function which is independent of $K_s$. It can be calculated\cite{dror} using 
the exact matrix elements available for $K_s=1/2$, with the result
\be
f_s(x)=c\left(1-\frac{x}{\sqrt{1+x^2}}\right) \; ,
\label{eq:fs}
\ee
where $c$ is a numerical constant.

The above extends the
earlier results of Voit\cite{voit} and Wiegmann.\cite{wiegmann} In
particular, the analytic structure (as a function of $k$ and $\omega$) of the
one soliton contribution to Eq. (\ref{eq:Gs}) reproduces that found in earlier
work.  The explicit discussion of the nonanalyticities due at the 
three soliton threshold is new, although fairly obvious; 
more muted singularities occur at the five and higher multisoliton 
thresholds, which we will not
discuss explicitly. The specific expression in Eq. (\ref{eq:Zs}) is, to
the best of our knowledge, new, and is the most important feature of this
result for the purposes of the present paper.

The charge pieces of both response functions are unaffected by the opening 
of the spin gap. Consequently, $\S$ and $G^<$ have  power law features 
(which can be a peak or a shoulder depending on $K_c$) at
$\omega= 2E_s(k/2) +{\cal O}(T)$ and $\omega= E_s(k)+{\cal O}(T)$, 
respectively, with a shape and temperature dependence, both readily computed,
determined by the still gapless charge density fluctuations.  
For example, 
we can evaluate the spectral function explicitly\cite{dror} at $T=0$ 
in the limit $v_s/v_c\rightarrow 0$ (and for arbitrary $\omega<3E_s(k/3)$),
or when $|\omega-\Delta_s|\ll \Delta_s$ (for arbitrary $v_s/v_c$ ):
\ba
\label{gkw}
\nonumber
G^<(k,\omega)&=&\frac{1}{4}
\frac{B(\gamma_c,\gamma_c+\frac{1}{2})}{\Gamma(\gamma_c)
\Gamma(\gamma_c+\frac{1}{2})}\left(\frac{\Lambda v_c}{2}\right)
^{-\frac{1}{2}-2\gamma_c} \\
&\times& Z_s(k)[\omega-E_s(k)]^{2\gamma_c-\frac{1}{2}}
\Theta[\omega-E_s(k)] \; .
\label{eq:gspingap}
\ea
Here $B(x,y)$ is the beta function.
Again, the fact that these excitations are not 
quasiparticles is reflected in the fact that, even where peaks in 
the spectral function occur, they do not narrow indefinitely as 
$T\rightarrow 0$.

In the presence of a spin gap, the spin contribution to the 
long distance behavior of the 
superconducting susceptibility is a constant;
\ba
\tilde \chi(x,t)&\sim & |\langle U_{s}^{2}\rangle |^{2}
\langle e^{i\sqrt{2\pi}\theta_{c}(x,t)}e^{-i\sqrt{2\pi}\theta_{c}
(0,0)}\rangle \nonumber 
\\
&\sim &(\Lambda\xi_s)^{-K_s}
\langle e^{i\sqrt{2\pi}\theta_{c}(x,t)}e^{-i\sqrt{2\pi}\theta_{c}(0,0)}
\rangle \; .
\label{eq:chitilde}
\ea
From this, one sees that, within a chain, one can identify 
$(\Lambda\xi_s)^{-K_s /2}$ as the ``amplitude'' 
and $\sqrt{2\pi}\theta_{c}$ as the ``phase'' of the order 
parameter.\cite{spingap}

\section{ Low Temperature:  The 3d Superconducting State} 
\label{sec:lowt}

For temperatures of order $T_c$ and below, interchain couplings cannot 
be ignored. Single particle hopping and all magnetic couplings are 
irrelevant by virtue of the preexisting spin gap. For $K_c > 1/2$, the 
Josephson coupling is perturbatively 
relevant, but for $K_c<1$, the $2k_F$ CDW coupling is more relevant.
For the simplest realizations of the 1DEG, $K_c <1$ corresponds to repulsive
interactions between charges.  However, we have shown\cite{nature,spingap} 
that for fluctuating or meandering stripes,
such as occur in the high temperature superconductors, the CDW coupling gets
dephased, so that the Josephson coupling is the most relevant, even when 
$1/2<K_c<1$.

Since we are interested in the onset of 
superconductivity, we consider the case in which
the Josephson coupling between chains is more relevant.
The pair tunnelling interaction between chains, which appeared in Eq. 
(\ref{eq:H}),
can be simply
bosonized:
\be
{H}_{J}=-
J_{SC}\sum_{<i,j>} \int dx [\hat \Delta_i^{\dagger}
\hat \Delta_j + {\rm H.C.}] \; ,
\label{eq:HJ}
\ee
where the pair creation operator on chain number $j$ is
\ba
\hat \Delta^{\dagger}(x,t)&=&
\psi_{1,\uparrow}^{\dagger}\psi_{-1,\downarrow}^{\dagger}+
\psi_{-1,\uparrow}^{\dagger}\psi_{1,\downarrow}^{\dagger} \nonumber \\
&\propto& \cos(\sqrt{2\pi}\phi_s)\exp(i\sqrt{2\pi}\theta_c) \; ,
\label{eq:pair}
\ea
and we have left the chain index implicit.  

Since the state below $T_c$ has long range order, and since we assume that the
coupling between chains is weak, it is reasonable to treat it
in mean field approximation,\cite{scalapino} although we continue to treat the
one-dimensional fluctuations exactly.  Thus, rather than considering a full
three-dimensional problem, we consider the effective  single chain problem 
defined by the Hamiltonian
\be
{\cal H}={\cal H}_s+{\cal H}_c - {\cal J} \cos(\sqrt{2\pi}\phi_s)
\cos(\sqrt{2\pi}\theta_c) \; ,
\label{eq:H4}
\ee
where ${\cal J}$ is related to the pair tunnelling amplitude by the mean field
relation
\be
{\cal J}=z J_{SC} (\Lambda/\pi)^2\langle \cos(\sqrt{2\pi}\phi_s)
\cos(\sqrt{2\pi}\theta_c)\rangle \; ,
\ee
where $z$ is the number of nearest neighbor chains. [Since the average of 
$\cos(\sqrt{2\pi}\phi_s)\sin(\sqrt{2\pi}\theta_c)$ vanishes, no sine term 
appears in the effective Hamiltonian (\ref{eq:H4}).] Note that the
pair hopping term in Eq. (\ref{eq:H4})
couples charge and spin, as is characteristic of higher dimensional 
couplings.

The mean field
approximation is exact in the limit of large $z$ 
and small $zJ_{SC}$.  In three dimensions, this
mean field approximation will produce some errors in the critical regime
in the vicinity of
$T_c$, but because of the long correlation length along the chain just 
above $T_c$, the critical region is always small for small $J_{SC}$, and  
well below $T_c$, this approximation is safe.\cite{meanfield}  

Because of the presence of relevant cosine terms, there are superselection 
rules which divide Hilbert space into various soliton sectors. The soliton 
sectors are specified by two integrals:
\ba
N_s= && \sqrt{2/\pi}\int_{-\infty}^{\infty} dx \partial_x\phi_s \\ 
= && \sqrt{2/\pi}\,[\phi_s(\infty)-\phi_s(-\infty)]
=2\int dx S_z \; , \nonumber
\ea
and
\be
\label{eq:Nc}
N_c=\sqrt{2/\pi}\int_{-\infty}^{\infty} dx \partial_x\theta_c 
=  \sqrt{2/\pi}\,[\theta_c(\infty)-\theta_c(-\infty)] \; . \nonumber
\ee

$N_s$
is simply the number of spin solitons minus the number of 
antisolitons or the total value of $S_z$ in units of $\hbar/2$.  
The interpretation of $N_c$ is a bit more
subtle. Since we are looking at a superconducting state, the electrostatic 
charge of a quasiparticle is not defined.
\cite{rokhsarandme,goldhaberandme,nayak}
However, $N_c$ is a conserved ``chirality''  equal to the number of 
right moving minus the number of left moving electrons, 
so that we can still interpret $eN_c$ as a sort of quasiparticle 
``charge'';  it represents the coupling of the quasiparticles to a 
magnetic flux.\cite{astute,rokhsarandme,goldhaberandme}  

The presence of the $\cos(\sqrt{8\pi}\phi_s)$ term in the single chain
Hamiltonian results in the quantization of $N_s$ in integer units.  
The presence of the
$\cos(\sqrt{2\pi}\phi_s)\cos(\sqrt{2\pi}\theta_c)$ term in
${\cal H}$ results in the quantization condition that $N_s+N_c$ be an even
integer!  Physically, this means that excitations can have spin $\hbar$ and
charge 0 ($N_s=2$ and $N_c=0$), spin 0 and charge $2$ ($N_s=0$ and $N_c=2$),
spin $\hbar/2$ and charge $1$ ($N_s=1$ and $N_c=1$), etc., but that all the
exotic quantum numbers of the soliton excitations of the isolated 1DEG are
killed.  
Formally, the addition of the pair hopping term to the Hamiltonian of the 1DEG
leads to a confinement phenomenon. Along the entire segment of chain between
two spatially separated
$\pm \sqrt{\pi/2}$ solitons, there is a change in sign of the pair hopping
term.  (See Eq. (\ref{eq:pair}).)
This leads to an energy which grows linearly with the separation $x$
between solitons, $\sim {\cal J} |x|$, regardless of whether  they are
charge or spin solitons or antisolitons.

The importance of this observation becomes clear when we study the operators 
in whose correlation functions we are interested.  Since
\be
e^{i\sqrt{\pi/2}\theta_s(x)}  \phi_s(y)e^{-i\sqrt{\pi/2}\theta_s(x)} 
= \phi_s(y) -\sqrt{\pi/2}\, \Theta(y-x) \; , 
\ee 
and 
\be
e^{i\sqrt{\pi/2}\phi_c(x)}  \theta_c(y)e^{-i\sqrt{\pi/2}\phi_c(x)} 
= \theta_c(y) +\sqrt{\pi/2}\, \Theta(x-y) \; ,
\ee 
it is clear that the fermion annihilation operator 
$\Psi_{-1,\uparrow}$ creates a 
spin antisoliton and a charge antisoliton, while the $2k_F$ piece of the 
spin-raising operator, ${S}_{2k_F}^+$, creates a pair of spin solitons 
and a pair of charge antisolitons.  
Both these combinations decay into a set of free solitons in the
absence of the interchain coupling, but in its presence, the former becomes a 
bound state, and the latter a resonant state.  
Thus, $G^<$ 
develops a coherent piece with a well defined dispersion relation as
superconducting phase coherence between chains occurs.  $\S$ develops
a resonant peak at a temperature well below $T_{c}$.

\subsection{Zero spin soliton sector}
For the case in which the spin gap $\Delta_s$ of the isolated chain is
large compared to the interchain coupling, the fluctuations of the spin 
field are high energy (fast)
compared to any charge fluctuations, and indeed only slightly affected by the
onset of superconducting order.  In this limit, 
the eigenstates can be treated in the adiabatic approximation. 

In the ground state ($N_s=0$) sector, the spin field fluctuations 
are little affected
by ${\cal H}_{J}$;  all spin correlations can thus be computed as 
in the previous
section.  Moreover, because of the spin gap, so long as $T\ll\Delta_s$, 
the spin fields
can be approximated by their ground state.  For computing the charge part 
of the wave function, we can replace the operator 
$\cos(\sqrt{2\pi}\phi_s)$ in ${\cal H}_{J}$ by its expectation value 
at zero temperature in the decoupled ground state,
\be
\cos(\sqrt{2\pi}\phi_s) \rightarrow {\langle\cos(\sqrt{2\pi}\phi_s)\rangle}_o 
\equiv {\cal C}_s\sim (\Lambda\xi_s)^{-\frac{K_s}{2}} \; ,
\ee
where the subscript $``o"$ refers to the expectation value in the 
ensemble with 
$J_{SC}$ set equal to zero, (see also Ref. \ref{ref:LukZam}).  
This leaves us with a sine Gordon equation for 
the charge degrees of freedom, with potential
\be 
{\cal J}{\cal C}_s \cos(\sqrt{2\pi}\theta_c) \; .
\ee
Again, we solve this problem
by refermionizing
\be
\Psi_{c,\lambda}^{\dagger}= {\cal 
N}_{c}\exp[i\sqrt{\pi/2}(\theta_c-2\lambda \phi_{c} )] \; .
\ee 
The refermionized form of the Hamiltonian is
\ba
{\cal H}_{c}= && i {\tilde v_{c}}[\Psi_{c,-1}^{\dagger}\partial_{x}\Psi_{c,-1}
-\Psi_{c,1}^{\dagger}\partial_{x}\Psi_{c,1}] \nonumber \\
&& - \tilde \Delta_{c}
[\Psi_{c,1}^{\dagger}\Psi_{c,-1}^{\dagger}+{\rm H.c.}] \nonumber \\ && +
g_c\Psi_{c,1}^{\dagger}\Psi_{c,-1}^{\dagger}\Psi_{c,-1}\Psi_{c,1} \; ,
\label{eq:Hc}
\ea
where 
\begin{eqnarray}
\nonumber
\tilde v_c&=&v_c\left(\frac{1}{4K_c}+K_c\right) \; , \\
\nonumber
\tilde \Delta_c&=&{\pi{\cal J}{\cal C}_s \over \Lambda} \; , \\
g_c&=&2\pi v_c\left(\frac{1}{4K_c}-K_c\right) \; .
\end{eqnarray}
Since $N_s=0$, the superselection rule implies 
$N_c=2m$, which upon refermionization is
simply the condition:
\be
-\sum_{\lambda}\lambda \int dx 
[\Psi_{c,\lambda}^{\dagger}\Psi_{c,\lambda}]=N_{c}/2=m \; .
\label{eq:fermionnumber}
\ee

It is also interesting to note that the superconducting pair creation 
operator can be expressed in an
intuitively appealing form in terms of charge soliton creation operators 
\be
\hat\Delta^\dagger\propto \cos(\sqrt{2\pi}\phi_s)
\Psi_{c,1}^{\dagger}\Psi_{c,-1}^{\dagger} \; .
\ee
Recall that here the charge solitons are spinless fermions.  This
expression emphasizes\cite{spingap} the fact that spin gap formation, which is
associated with the quenching of the fluctuations of the spin density phase,
$\phi_s$, can also be identified with the growth of the {\it amplitude} of the
superconducting order parameter.  While the charge solitons clearly also 
make a contribution to the amplitude of the order parameter,  
the phase of the order parameter comes entirely from the charge.

For $K_c=1/2$, just as for
the Luther-Emery point for the spin fields, the refermionized Hamiltonian 
for the charged excitations is noninteracting and massive (gapped), and 
$\Delta_c = \tilde \Delta_c$.
In computing the asymptotic form of correlations we will set $K_c=1/2$.
We can now readily compute the expectation value of
the pair hopping term so as to relate two physically 
important quantities: the excitation energy scale, $\Delta_c$, 
and the interchain portion of the internal energy 
\ba
\Delta_{c}\langle \Psi_{c,1}^{\dagger}\Psi_{c,-1}^{\dagger}+{\rm H.c.}\rangle 
&& ={\cal J}\langle\cos(\sqrt{2\pi}\phi_s)\cos(\sqrt{2\pi }\theta_c)
\rangle \nonumber \\
&& =(\Delta_c/\pi \xi_c) u_0(\Delta_c,T) \; ,
\label{eq:HSC}
\ea
where $\xi_c=v_c/\Delta_c$ is the charge correlation length. 
Equation (\ref{eq:HSC}) has the form of a BCS gap equation with
\be
u_0(\Delta_c,T)=\int_0^{v_{c}\Lambda} dx 
\frac {1} {\sqrt{x^2 +\Delta_c^2}} \tanh\left( {1 \over 2 T} 
\sqrt{x^2 +\Delta_c^2}\right) \; ,
\label{eq:u0}
\ee
where the mean field relation for $\Delta_c(T)$ is
\be
u_0(\Delta_c,T)={\pi v_c \over
 zJ_{SC}{\cal C}_s^2} \; .
\label{eq:gapeq}
\ee 
Consequently we find the familiar BCS relations 
\be
T_c = 0.57 \Delta_c(0) \; ,
\ee
\be
\Delta_c(0)=2v_c\Lambda\exp[-\pi v_c /
 zJ_{SC}{\cal C}_s^2] \; ,
\label{eq:deltac}
\ee 
\be
\label{apptb}
\Delta_c(T)\approx 1.74\Delta_c(0)\sqrt{1-T/T_c} \;\;\;\;\;\; 
{\rm for} \; T\approx T_c \; .  
\ee
In general, the actual form of $\Delta_c(0)$ in terms of $J_{SC}$ and 
${\cal C}_s$ is modified according to the microscopic value of $K_c$. 

The transverse superconducting phase stiffness $\kappa_{\perp}$ (proportional 
to the superfluid density) is 
\be
\kappa_{\perp}=2\pi a\langle H_{J}\rangle \; ,
\label{eq:kappaperp} 
\ee
where $d$ is the spacing between chains and $\langle H_{J}\rangle$ 
is given in Eq. (\ref{eq:HSC}).  Thus at zero temperature 
$\kappa_{\perp}\sim T_c^2/v_c$. As is shown in Appendix B, 
for a system with equal areas of domains 
in which the stripes run along the $x$ and $y$ directions, the macroscopic 
phase stiffness is equal to the geometric mean of the 
superfluid density in the directions parallel and perpendicular to the
chains, $\bar \kappa=\sqrt{\kappa_{\parallel}\kappa_{\perp}}$.  
Since the phase stiffness along the chains is simply 
$\kappa_{\parallel}=v_cK_c$, it follows that $\bar \kappa(T=0)$   
is (up to logarithmic corrections coming from $u_0$) 
simply proportional to
$T_{c}$.   This is a microscopic realization of a more general 
phenomenon which occurs in systems with low superfluid density;\cite{phase} 
it is phase ordering, as opposed to pairing, which determines $T_{c}$.  
In a future publication\cite{phase2}, we will study the effects of quantum 
and thermal phase fluctuations on
the evolution of the superfluid density of a 
quasi one-dimensional superconductor.

With little additional effort, we can study the pair field 
susceptibility $\chi$ at energies small compared to $2\Delta_s$.
In this low energy limit, as in Eq. (\ref{eq:chitilde}), 
we can replace the spin operators in $\chi$ by their ground state 
expectation values.  

The charge part of $\tilde\chi$ can be expressed in terms of 
the charge fermion fields
\be
\tilde\chi_c(x,t)=\langle\Psi_{c,1}^{\dagger}(x,t)\Psi_{c,-1}^{\dagger}(x,t)
\Psi_{c,-1}(0,0)\Psi_{c,1}(0,0)\rangle \; .
\ee
At the free charge fermion point ($K_c=1/2$) the corresponding spectral 
function is readily evaluated, for $T=0$ and $\omega\ll 2\Delta_s$, 
with the result
\ba
\nonumber
\chi(k,\omega)&=&\left(\frac{{\cal C}_s u_0}{\xi_c}\right)^2\delta(k)
\delta(\omega) \\
\nonumber
&+& \frac{{\cal C}_s^2[\omega^2-4E_c^2(k/2)+2\Delta_c^2]}{4v_c^2|q_1 E_c(q_2)
-q_2 E_c(q_1)|}\Theta[\omega-2E_c(k/2)] \; , \\
\ea
where $E_c(k)$ and $q_{1,2}$ are the analogs of Eqs. 
(\ref{es1def},\ref{q1q2s})  
with $\Delta_c$ substituted for $\Delta_s$ and $v_c$ for $v_s$.

Away from the Luther-Emery point, if
$g_c >0$ ($K_c <1/2$), the two solitons repel, 
and hence the effect of $g_c$ can be
ignored, but for $g_c <0$ ($K_c> 1/2$), 
there is an attractive interaction between the two solitons and hence,
this being after all a one dimensional problem, they form a bound state.  
This will slightly modify the expression for $\chi$.

\subsection{The one hole sector}

In the one soliton sector of the spin Hamiltonian, the adiabatic approximation
requires reexamination.  While for the most part, the spin modes are fast
compared to the charge modes, the Goldstone mode (translation mode of the spin
soliton) is slow compared to all other modes, and so must be treated in the
inverse adiabatic approximation.  Thus, we consider the charge Hamiltonian
with a spin antisoliton at fixed position, $R_s$. The pair tunnelling term is 
then  
\be
{\cal J}{\cal C}_s {\rm sign}(x-R_s) \cos(\sqrt{2\pi} \theta_c) \; ,
\ee
where we have used the fact that $\xi_s=v_s/\Delta_s$ (which
characterizes the width of the spin soliton) is small compared to the charge
correlation length,
$\xi_c=v_c/\Delta_c$, to approximate the profile of the spin
soliton by a step function.  Upon refermionization, the charge Hamiltonian is
still of the same form as Eq. (\ref{eq:Hc}) with the term proportional to
$\tilde\Delta_c$ replaced by
\be
\tilde\Delta_c \rightarrow -\tilde\Delta_c {\rm sign}(x-R_s) \; .
\label{eq:Hckink}
\ee
For $K_c=1/2$, upon the canonical transformation, 
\be
\psi_{c,-1}=\Psi_{c,-1}^{\dagger} \; , \;\;\;\;\;\;\;\; 
\psi_{c,1}=\Psi_{c,1} \; ,
\ee
the charge soliton Hamiltonian is of the same form\cite{chx} as
the fermionic Hamiltonian of a commensurability 2 Peierls insulator, 
``polyacetylene'', in the presence of a topological soliton.  
As is well known,\cite{chx} there is an index theorem
that implies the existence of a zero energy bound state associated
with the soliton, the famous ``midgap state'' or ``zero mode''.  All
other fermionic states have energies greater than or equal to
$\Delta_c$.  Importantly, since in this sector
$N_s=-1$, the superselection rule $N_c=2m+1$, requires that the fermion 
number is half integer!
\be
-\sum_{\lambda}\int dx 
:[\psi_{c,\lambda}^{\dagger}\psi_{c,\lambda}]:\, =N_{c}/2=m+1/2 \; .
\ee
This is essential, since with the midgap state occupied the
fermion number is\cite{chx} $+1/2$, while with it empty
the fermion number is $-1/2$.  The midgap state is
associated with the bound state of the spin and charge antisolitons.

To compute the charge contribution to the soliton creation energy we need to
evaluate the difference between the ground state energies of the charge
Hamiltonian in the presence and absence of a kink.  We have done this by
taking the limit of vanishing soliton width of a general expression of
Takayama, Lin-Liu, and Maki,\cite{tlm}  
(and dividing by 2 for the spinless case).
The resulting soliton creation energy is just
$\Delta_c/2$ ;
in other words, the rest energy of the electron, {\it i.e.} 
the bound state of a spin soliton and a charge soliton, is 
\be
\Delta_0=\Delta_s+\Delta_c/2\approx \Delta_s \; .
\label{eq:delta0}
\ee

From this discussion, we can immediately conclude 
that for $T\ll T_c \ll \Delta_s$, the
one hole spectral function has a coherent piece and a multiparticle
incoherent piece,
\be
G^<(k,\omega)=Z(k)\delta[\omega-{\cal E}(k)] + 
G^{(multi)}(k,\omega) \; ,
\label{eq:G}
\ee
where
\be
{\cal E}(k)=\sqrt{v_s^2k^2+\Delta_0^2} \; .
\ee
This follows from the fact that the bound state of a spin soliton and
a charge soliton has the same quantum numbers as a hole.  
The multiparticle piece\cite{notagap}
has a threshold slightly above the single hole threshold at
$\omega={\cal E}(k)+2\Delta_c$.

The overlap factor, $Z(k)$, 
contains factors from
both the spin  and the charge parts of the wavefunction;  
so long as $k\xi_s \ll 1$, $Z(k)=Z_c(k)Z_s(0)$
where $Z_s(0)$ depends on the
spin correlation length as in  Eq. (\ref{eq:Zs}), 
and $Z_c(k)$ contains all remaining contributions.
We can obtain a scaling form for $Z_c$ using 
the same method of analysis employed
previously for $Z_s$.  
Specifically, at $T=0$ in the absence of interchain coupling, and for
$\omega \ll 3\Delta_s$ and $|k\xi_s| \ll 1$, $G^{<}$ is given by the 
expression in Eq. (\ref{eq:gspingap}).
Since the opening of a charge gap does not affect the high 
energy physics, the dependence of $G^<$ on $\Lambda$ is unaffected 
by the interchain coupling.  Indeed, so long as $\Delta_c\ll \Delta_s$, 
the dependence of
$G^<$ on $\Delta_s$ is likewise unchanged.  Thus, by
dimensional analysis, it follows that
\be
Z(k)=Z_s(0)(\Lambda\xi_c)^{-\frac{1}{2}-2\gamma_c}A_{\gamma_c}
\tilde f(k\xi_c) \; ,
\label{eq:zofk}
\ee
where $\tilde f$ is a scaling function and 
\be
A_{\gamma_c}= 
{B(\gamma_c,\gamma_c+1/2)\over \Gamma(\gamma_c)\Gamma(\gamma_c+1/2)} \; .
\ee
Unfortunately, we do not have 
exact results from which to
compute $\tilde f(x)$ explicitly, but there is no reason to expect 
it to have any very interesting
behavior for small $x$.

At temperatures between $T=0$ and $T=T_c$, the same arguments lead to a simple
approximate
expression for the spectral function.  Specifically, the principal 
temperature dependence
comes from $\Delta_c$ which is a decreasing
function of $T$.  At mean field level, 
the temperature dependence of $\Delta_c$ can be
computed from Eq. (\ref{eq:u0}). In particular it vanishes at $T_c$ according 
to  Eq. (\ref{apptb}).  
Since fluctuation effects produce superconducting
correlations between neighboring chains at temperatures above 
$T_c$, this simple
mean field behavior will be somewhat rounded, 
but the qualitative point that $\Delta_c$
becomes small at temperatures above $T_c$ is quite robust.  

Consequently, the quasiparticle weight, $Z$, which is proportional 
to $\Delta_c^{2\gamma_c+\frac 1 2}$, is a strongly decreasing function of
$T$ which vanishes in the neighborhood of $T_c$. The quasiparticle gap,
$\Delta_0$, on the other hand, is only weakly temperature dependent, 
dropping from its maximum value $\Delta_0=\Delta_s+\frac 1 2 
\Delta_c(0)$ at $T=0$ 
to $\Delta_0=\Delta_s$ in the neighborhood of $T_c$. Scattering off thermal 
excitations will, of course, induce a finite lifetime for the quasiparticle 
at finite temperatures.   

Neither a charge soliton nor a spin soliton can hop from one chain to the 
next, but a hole can.  The problem of the transverse dispersion of the 
coherent peak in the single hole spectral function is
addressed in Appendix A. Not surprisingly, we find that the effective 
interchain hopping matrix element, $t_{\perp}$, is replaced by an effective 
interchain hopping matrix element,
\be
t_{\perp}^{eff}=Z(k) t_{\perp} \; .
\label{eq:teff}
\ee
Thus, the dispersion of the coherent peak transverse to the chain 
direction is an independent measure of the degree of interchain coherence.

\subsection{The two spin soliton sector}

To compute $\S$, we need to study states in the $N_s=2$
sector. Interestingly, (in contrast to the case of an
ordered CDW) in a quasi one-dimensional superconductor, the $2k_F$ spin 
density wave operator also creates two charge antisolitons:  $N_c=-2$.  
Again, for the most part, the spin fluctuations are fast and  high
energy compared to the scale of the charge fluctuations, and can thus  
be treated in the adiabatic approximation - indeed,
they are little affected  by the presence of the interchain 
Josephson coupling. However, there are two low frequency modes associated 
with the soliton translational degrees of freedom, 
which must be treated in the antiadiabatic approximation.  
Consequently, we obtain an effective 
Schr\"odinger equation governing the
center of mass motion of the two spin solitons:
\be
H^{eff}\approx 2\Delta_s -{1\over 2M^*}\sum_{j=1}^2 {\partial^2\over 
\partial x_j^2} + V(x_1-x_2) \; ,
\label{eq:Heff}
\ee
where $x_j$ is the position of soliton $j$,
\be
M^*= \Delta_s/v_s^2 \; ,
\label{eq:Mstar}
\ee
and $V$ is the adiabatic spin soliton potential, 
obtained by integrating out the (relatively fast) fluctuations
of the charge degrees of freedom.

To compute $V(R)$, we again rely on the analogy between the refermionized 
version of the charge part of the Hamiltonian and solitons in
polyacetylene.\cite{chx,ssh,tlm} 
In this case, $V(R)$ is recognized as the difference in  
the ground state energy of a massive Dirac fermion in the presence 
and absence of a pair of
zero width solitons separated by a distance $R$, {\it i.e} the Hamiltonian in 
Eq. (\ref{eq:Hc}) with
\be
\tilde\Delta_c\rightarrow \tilde\Delta_c {\rm sign}(4x^2-R^2) \; .
\ee  
Since $2N_c=-2$, this energy difference is to be computed in the 
fermion number $-1$ sector.

From the results in the previous section, it follows that
\be
V(R)\rightarrow \Delta_c \ \ {\rm as} \ \ R\rightarrow \infty \; ,
\ee
since in this limit, the two solitons are noninteracting, 
and reduce to the solution discussed in the
previous section. Similarly, since for $R=0$, the energy approaches that
of the uniform system with fermion number $-1$,
\be
V(R)\rightarrow \Delta_c \ \ {\rm as} \ \ R\rightarrow 0 \; .
\ee
Moreover, from simple scaling, it is clear that
\be
V(R)=\Delta_c [1+v(R/\xi_c)] \; ,
\ee
where $v(x)$ is independent of the magnitude of $\Delta_c$ and
$v(x)\rightarrow 0$ for $x\rightarrow 0$ and $x\rightarrow \infty$.
For intermediate $R/\xi_c$, we have been unable to obtain an analytic
expression for $v$, although it is easily derived numerically,
as described in Appendix C, with the result shown in Fig. 1.  
As can be seen, $v(x)$ rises from 0 to a gentle maximum at 
$x\approx 0.3$ where  $v(0.3)\approx 0.2$, and then drops
exponentially back to zero at large separation.  

What this means is that there is no true bound state 
in the spin one excitation spectrum.
The spin one excitations, even in the superconducting state, 
are always unstable to decay into a pair of far separated spin 1/2 
quasiparticles.  However, near the threshold energy, 
$\omega=2\Delta_s+\Delta_c$, there is a nearly bound (resonant)
state with a lifetime which is exponentially long.  
Treating Eq. (\ref{eq:Heff}) in the WKB approximation,
we see that the decay rate of the resonant state is
\be
\Gamma \sim \exp\big[-B(v_c/v_s)\sqrt{\Delta_s/\Delta_c}\  \big] \; ,
\ee
where $B$ is a constant of order 1. 

\begin{figure} 
\narrowtext
\setlength{\unitlength}{1in}
\begin{picture}(3.2,3.0)(0,-1)
\put(-0.25,-0.8){\psfig{figure=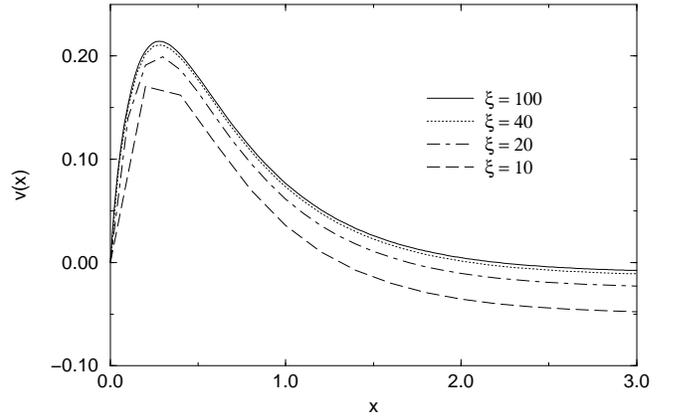,width=3.7in}}
\end{picture}
\caption{The scaled version of the adiabatic potential, $v(x)$, 
defined in Eq. (67), 
computed from the SSH model in a system of size 3000 sites with open
boundary conditions.  The different curves are for different magnitudes
of dimerization corresponding to a coherence length of the indicated
magnitude in units of the lattice constant.}
\label{fig1}
\end{figure}

Using the fact that ${\cal S}\sim\Lambda^{-\frac{1}{K_c}}$ in the 
absence of a charge gap and utilizing the same scaling arguments applied 
previously to the coherent piece of $G^<$, it is easy to see that the 
weight associated with this resonant state is proportional to 
$\Delta_c^{\frac{1}{K_c}-1}$. However, because the barrier height is 
small compared to $\Delta_c$, the thermal decay of the resonant bound 
state will become large, due to activation over the barrier, at a 
temperature well below $T_c$.

\subsection{ The ``BCS-like case'':  no preexisting spin gap}

When there is no spin gap on the isolated chain and there are 
repulsive interactions in the charge sector, the interchain Josephson 
coupling is perturbatively irrelevant.  Thus, the usual case for a 
quasi one-dimensional superconductor 
is the already analyzed case with a preexisting 
spin gap.  However, it is worthwhile considering the case (with 
$K_{c}<1$) in which both the spin gap and the superconducting 
coherence are induced by a relevant interchain Josephson coupling.  
This case, even though quasi one-dimensional, is much more akin to 
the usual BCS limit, in that there is a single gap scale in the 
problem, and pairing (gap formation) and superconducting coherence 
occur at the same temperature and with roughly the same energy 
scale.  This case has been analyzed extensively in the
literature.\cite{brazovskii,bourbonais,schulz}
It should be noted, however, that here, too, since the ``normal'' state
is a non-Fermi liquid, the coherent piece of all spectral functions
will be strongly temperature dependent below $T_c$, and vanish
in the neighborhood of $T_c$.

\section{Summary and Implications for Experiment}
\label{sec:summary}

In this paper, we have obtained explicit and detailed results for the 
properties of the superconducting state of a quasi one-dimensional 
superconductor.  We have studied this problem as a quantum critical 
phenomenon, in which the quantum critical point is reached in the 1d 
limit of no interchain coupling, and hence we have treated the 
interchain Josephson coupling as a small parameter.  In particular, we 
expect (as discussed below) that the results are pertinent to 
underdoped and optimally 
doped high temperature superconductors, where self-organized stripe 
structures render the system locally quasi one-dimensional.  

It is often argued that, even in fairly exotic circumstances, and even 
when the normal state is a non-Fermi liquid, the superconducting state 
itself is fairly 
conventional and BCS-like.  We have shown that there are a number of 
ways in which this expectation is violated.  In the first place, there 
are two ``gap'' scales, $\Delta_{s}\gg \Delta_{c}$, whereas in a BCS 
superconductor there is one, $\Delta_{0}$, and correspondingly two 
correlation lengths, $\xi_{s}$ and $\xi_{c}$, in place of the one, 
$\xi_{0}$, of BCS theory.  However, both gaps are, in a very 
real sense, superconducting gaps:  $\Delta_s$ is associated with spin 
pairing ({\em i.e.} a nonzero value of 
$\langle\Psi_{s,1}^{\dagger}\Psi_{s,-1}\rangle $) 
and the existence of a local amplitude of the order parameter.
$\Delta_c$ is a measure of 
interchain phase coherence. In the case in which
there is a  preexisting spin gap on the isolated chain, 
$\xi_{s}$ remains finite 
at the quantum critical point, whereas $\xi_{c}$ diverges.   The same 
holds true in the superconducting phase, a bit away from the quantum 
critical point, where $\xi_{c}$ diverges at $T_{c}$, while $\xi_{s}$ 
remains finite.\cite{notagap}  

It is perhaps worth noting that many of these novel aspects of the
superconducting state are considerably more general than the particular 
model we have solved.  Indeed, recently, Lee\cite{dunghai} derived similar 
results from the gauge theory formulation of a flux phase to superconductor 
transition.  While this derivation presupposes rather different seeming 
microscopic physics, it does build in the doped insulator character of 
the superconducting state, which is the essential feature of the results.  
Likewise, many features we have discussed here bear a close resemblance 
to the dimensional crossover from a conjectured 2d non-Fermi liquid to a 3d
superconductor envisaged in the context of the inter-layer tunnelling 
mechanism of high temperature superconductivity\cite{ilt}.

\subsection{Summary of Results}

For the benefit of the reader who skipped the technical exposition, 
we begin by summarizing our most important results.  
We consider here the case in which there is a
preexisting spin gap, $\Delta_s/2 \gg T_c$, on an isolated chain, 
and we focus on the effects of the
interchain Josephson coupling between stripes at lower energies.  

\subsubsection{Thermodynamic Effects}

The effect of the interchain Josephson coupling is to produce an 
interchain coherence scale,
$\Delta_c(T)$.  At mean field level, $\Delta_c(T)$ vanishes 
for any $T$ above $T_c$, and while
fluctuation effects will produce a small amount of rounding 
to this behavior, because of the
large coherence lengths along the chain the degree of rounding 
will always be small in the quasi-1D limit. It is the coherence scale 
that determines $T_c$, in the sense that 
\be
T_c\approx \Delta_c(0)/2 \ll \Delta_s/2 \; .
\ee 
($\Delta_c$ is expressed in terms of the strength of the Josephson 
tunnelling matrix elements in
Eq. (\ref{eq:deltac}).)  
The superfluid densities in the directions transverse and parallel to
the chain direction are, respectively,
\be
\kappa_{\perp}= 2a u_0\Delta_c^2/ v_c\; , \ \ \kappa_{\parallel}=v_cK_c \; ,
\ee
where $u_0$ is a constant (see Eq. (\ref{eq:u0})) which depends weakly on 
parameters, $d$ is the
spacing between chains, $v_c$ is the charge velocity, and $K_c$ is the 
charge Luttinger parameter.  
In two dimensions, if, on average, there is a 4-fold rotationally invariant
mixture of domains in which the
chains run along the x and y directions, respectively, the
macroscopic superfluid density is isotropic and given by
\be
\bar \kappa(T) = \sqrt{\kappa_{\perp}\kappa_{\parallel}} 
\sim \Delta_{c}(T) \; .
\ee

\subsubsection{Single Hole Spectral Function}

The common theme in the spectral functions is that all dependence on
the interchain coupling (and hence all important temperature dependences 
in the neighborhood of
$T_c$) are expressible in terms of the single coherence scale, 
$\Delta_c$.  Moreover, it is the
spectral weight of the coherent features in the spectrum, 
rather than their energies, which are
strongly temperature dependent!  
This is very different from the behavior of the spectral
functions near $T_c$ in a three dimensional BCS superconductor.

Characteristic shapes of the single hole spectral function
above and below $T_c$ are shown in Fig. 2.  
Above $T_c$, the single hole spectral function is a broad
incoherent peak.
Below $T_c$, there is a coherent delta-function piece and a
multiparticle continuum at higher energy,
\be
G^<(\vec k,\omega)=Z( k_{\parallel})\delta[\omega-{\cal E}(\vec k)] + 
G^{multi} \; ,
\ee
where
\be 
{\cal E}(\vec k)=\sqrt{v_s^2 k_\parallel^2+\Delta_0^2}+
2t_\perp Z(k_\parallel)\cos(k_\perp a)+... \; . 
\ee 
Here $k_F+k_\parallel$ and $k_\perp$ are, respectively, the components of 
the crystal momentum parallel and perpendicular to the chain direction. 

\begin{figure} 
\narrowtext
\setlength{\unitlength}{1in}
\begin{picture}(3.2,3.3)(0,-1.2)
\put(-0.25,-0.6){\psfig{figure=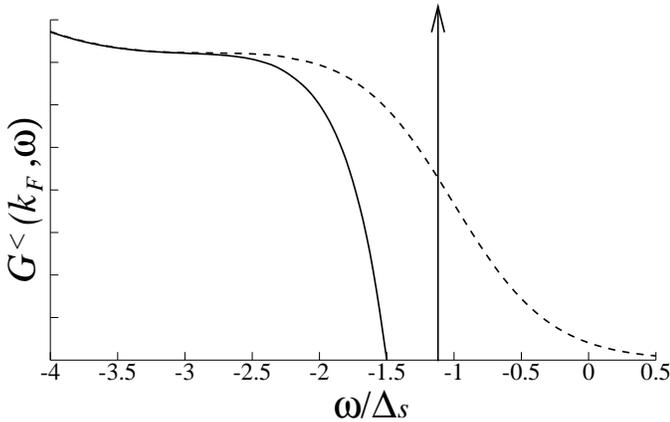,width=3.5in}}
\end{picture}
\caption{The temperature evolution of the spectral function. The dashed 
line depicts $G^<(k_F,\omega)$ at temperature $T=\Delta_s/3>T_c$, 
and is calculated using the parameters $\gamma_c=0.3$, $K_s=1/2$ and 
$v_s/v_c=0.2$. The solid line represents the spectral function at zero 
temperature. A coherent $\delta$-function peak onsets near $T_c$ 
at energy $\Delta_0=\Delta_s+\Delta_c(0)/2$. Here we assume 
$\Delta_s/\Delta_c(0)=5$. The multi-particle piece starts at a 
threshold $2\Delta_c(0)$ away from the coherent peak. The exact 
shape of the incoherent piece at $T=0$ is not calculated in the present work 
and is meant to be schematic.}
\label{fig2}
\end{figure}

The energy gap for the coherent peak is 
\begin{equation}
\Delta_0(T)=\Delta_s+\frac{1}{2}\Delta_c(T) \; ,
\end{equation}
and its spectral weight is given by
\be
Z(k)\sim [\Delta_c(T) ]^{2\gamma_c + \frac 1 2} \; .
\label{eq:Zcrit}
\ee 
Thus, $Z(k)$ (and with it the transverse bandwidth) is the most 
strongly temperature dependent feature of the spectral function.

The multiparticle incoherent piece $G^{multi}$ starts at a threshold
energy ${\cal E}(\vec k) + 2 \Delta_{c}(T)$.  
This is the origin of the gap between the coherent peak and the incoherent 
shoulder in Fig. 2. Various forms of damping, including phase 
fluctuations transverse to the stripes, will broaden this structure, 
leading to a peak-dip-shoulder 
form of the spectral function. However, the distance from the coherent 
peak to the dip should be proportional to $\Delta_c(T)$ and hence, at $T=0$,
to $T_c$.

\subsubsection{The Spin Response Function}
\label{sec:twospinon}

The spin response function is entirely a multiparticle continuum;  even below
$T_c$, we find that any spin-1 mode is unstable to decay into two spin 1/2
``quasiparticles''.  However, at low temperature, we find that there is
a spin-1 resonant state with an exponentially long lifetime near 
the threshold energy $2 \Delta_s + \Delta_c = 2\Delta_0$, with 
momentum $2 k_F$, where $k_F$ is the Fermi momentum on a stripe.  Even here,
because the barrier to decay is quantitatively small compared to $T_c$,
we expect that no sharp resonant state will appear in the spectrum
in the neighborhood of $T_c$.  Rather, it will appear as the temperature
falls below $T\approx 0.2\Delta_c(0)\approx 0.4T_c$.  

\subsection{Implications of Two Scales}

The existence of two scales in the superconducting state
appears in  different experiments in
fairly obvious ways:
 
{\bf 1)}  Since an electron has spin and charge, the gap measured in single 
particle spectroscopies, such as ARPES or 
tunnelling, is $\Delta_{0}=\Delta_{s} +(1/2) \Delta_{c}(T)$. (See Eq. 
(\ref{eq:delta0}).)  Manifestly, this gap scale 
decreases slightly with increasing temperature, but remains large, roughly 
$\Delta_{s}$, above $T_{c}$.  The gap scale $\Delta_s$ is 
unrelated to $T_{c}$, and moreover $\Delta_{s}(T=0) \gg 2T_{c}$, 
which physically is the 
statement that the onset of phase coherence, {\em not} pairing, is what 
determines $T_{c}$. Consequently, the zero-temperature superfluid 
density is a better predictor\cite{uemura,phase} of $T_{c}$ than 
$\Delta_{0}(T=0)$.  (See Eq. (\ref{eq:kappaperp}) and subsequent discussion.)  
Similarly, pure spin probes, such as NMR or neutron scattering, 
see a gap which is approximately $\Delta_{s}$ per spin 1/2.  
(See Eqs. (\ref{eq:Heff}).)

{\bf 2)}
Experiments involving singlet pairs of electrons, such as Andreev 
tunnelling, could exhibit\cite{notagap} an energy scale $\Delta_{c}$; 
a scale, moreover, which vanishes at (or near) $T_{c}$, and is related 
in magnitude to $T_{c}$ in a more or less familiar manner, 
$\Delta_{c}(0)/2 \sim T_{c}$.  More complicated
spectroscopies, such as SIS tunnelling ({\em e.g.} tunnelling across 
a break junction)  should reveal gap-like features with both energy 
scales, $\Delta_{s}$ and $\Delta_{c}$.  

{\bf 3)}  The existence of two correlation lengths implies that different 
measurements will find the order parameter magnitude depressed over 
distinct distances:  If an impurity destroys the
superconducting gap locally, the single-particle density of states, 
as determined, for instance,
with a scanning tunnelling microscope, will basically recover over a length
scale $\xi_{s}$ (although, subtle effects will persist out to a scale $\xi_c$).
By  contrast, the magnetic field strength
near the core  of a vortex, which otherwise
would diverge  logarithmically at short distances, is reduced inside a 
``core radius'' due to the fact that the superfluid density is depressed, 
({\it i.e.} there is a lower current density per unit phase gradient).  
Since this latter effect involves
only charge motion, the vortex core radius is of
order $\xi_{c}$.  This ``magnetic'' core radius is measured, in 
principle, in $\mu$SR.\cite{sonier}  

{\bf 4)}  The superconducting state
reflects the non-Fermi liquid character of the normal state in many 
ways, but it has a complex scalar order parameter 
as in a conventional (BCS) superconducting 
state.  This means that we {\it might} expect well-defined 
elementary excitations with the quantum numbers
\cite{rokhsarandme,goldhaberandme,nayak}
of the electron quasiparticle, as indeed we have found.   
However, in a conventional superconductor,
the quasiparticle  energy is shifted by the opening of the gap, and the 
lifetimes of all elementary excitations (as observed {\it e.g.} in ultrasonic 
attenuation) are strongly temperature dependent 
below $T_{c}$.  In the present case, it is the {\it spectral weight} 
associated with the elementary excitations 
which is strongly temperature dependent, not the 
lifetime or the energy.  Moreover, even as $T\rightarrow 0$, the 
quasiparticle weight remains small, in proportion to a positive power of 
the distance from the quantum critical point.  See Eq. (\ref{eq:Zcrit}).

\subsection{ Two Scales in the High Temperature Superconductors}

It has been noted,\cite{yamada,bourges} in the so-called ``Yamada plot,'' that
$T_c$ in underdoped high temperature superconductors is proportional 
to the observed 
incommensurability\cite{incom1,incom2} in the low energy spin structure 
factor.  The magnetic incommensurability, $\pi/d$,  is inversely 
related\cite{tranquada,oron} to the mean separation between charge 
``stripes,'' $d$.  Thus, the Yamada plot implies that $T_c$ is 
inversely proportional to the mean spacing between stripes;  As
the stripes become more separated, and the electronic structure becomes more 
one dimensional, $T_c \rightarrow 0$. This observation strongly supports the 
idea that the anomalous electronic properties of these materials reflect the 
properties of nearby phases of the 1DEG.  Indeed, 
many of the spectral features  
listed above have been observed, with various levels of  confidence, 
in experiments on the high temperature superconductors: 

{\bf 1)}  The best 
single particle spectra (ARPES and tunnelling) exist for {\BSCO} 
because it cleaves easily.  For underdoped and optimally doped materials, 
the single particle gap as measured by tunnelling and 
ARPES\cite{harris} is found to be 
large: $\Delta_{0}\ge 35$meV, in the ``flat-band'' region near the ${\bar M}$, 
or $(\pi,0)$ and $(0,\pi)$, points of the Brillouin zone.  
($\Delta_{0}/2T_{c}$ lies in the range 2 to 6. The ${\bar M}$ region is where 
the maximum of a d-wave superconducting gap is expected.) The gap 
persists\cite{persists} in some form or other to temperatures well above 
$T_{c}$.  Moreover, $\Delta_0$ increases with underdoping while $T_{c}$ 
decreases.  This gap is quite clearly a superconducting gap in that it has 
(at low $T$) the characteristic d-wave form\cite{lowTgap} expected of the 
superconducting gap, and it evolves\cite{evolves} 
smoothly with overdoping into a gap of only slightly smaller magnitude 
which opens, in a more conventional manner, in the neighborhood of 
$T_{c}$.  (We focus on the gap near the ${\bar M}$ point, especially, 
because there are both theoretical\cite{salkola2} and 
experimental\cite{shen2} reasons to think that the ``flat-band'' 
region is associated with states in ``stripes'' or ``fluctuating 
stripes.'')  Moreover, as discovered first by Uemura and coworkers,
\cite{uemura} $T_{c}$ is roughly proportional to the zero 
temperature superfluid 
density for underdoped materials, consistent with the notion\cite{phase} 
that it is phase ordering, not pairing, which determines $T_{c}$.

{\bf 2)}  Deutscher\cite{deutcher} has argued that the 
gap scale determined by low temperature Andreev tunnelling 
spectroscopy is considerably smaller than that determined from single 
particle tunnelling measurements in underdoped materials, while the 
two gap scales approach each other in overdoped materials.  
This issue is well worth revisiting in more detail.
The single particle gap scale is strongly apparent\cite{sis} in SIS  
tunnelling spectra - we do not know of any convincing analysis
which  reveals the smaller charge gap scale in such experiments.

{\bf 3)}  The vortex core radius has been measured with both 
scanning tunnelling microscopy\cite{renner,davis} (STM) 
and $\mu$SR.\cite{sonier}  
The $\mu$SR study  measures the magnetic field distribution in the material, 
and infers the core radius from the high-field cutoff of the distribution.  
For large applied 
fields ($B\sim 6T$),
both methods are in rough agreement that the core 
radius is about 15\AA.  However, the core radius deduced from the 
$\mu$SR measurements is strongly field dependent, so that at low fields 
($B\sim 0.5T$) it yields a core radius around 120\AA.  By contrast, 
preliminary evidence\cite{private} from STM experiments suggests that the core 
radius measured by that method is not strongly field dependent, so 
that, in low fields, the results of the two methods differ by almost an 
order of magnitude. However, there appear to be differences in
the STM results of different groups.\cite{davis}  Certainly,
the core radius inferred from STM studies\cite{ali,davis2} of the 
gap suppression in the vicinity of an impurity at zero magnetic field are
suggestive of a rather short coherence length.  
While the experimental results are, by no means, definitive, 
we would tentatively
like to explain the discrepancy between the STM and $\mu$SR results
at low field as evidence of the existence of two coherence lengths in
the superconducting state.

{\bf 4)} It has been realized for a long time that there are no sharp 
quasiparticle
features in the ARPES spectrum near the superconducting gap maximum 
(near the ${\bar M}$ point of the Brillouin zone) in the normal 
state, and it has been argued\cite{argued} that they disappear 
due to a lifetime catastrophe which occurs as the temperature is raised above
$T_{c}$.  Recent\cite{fedorov,shen} high-resolution 
ARPES measurements in optimally doped {\BSCO} have revealed 
a new picture of the emergence of these peaks.  Within experimental 
resolution, neither the energy nor the width of the 
peak changes as the temperature is raised from well below $T_{c}$ to 
slightly above $T_{c}$;  rather, it is the intensity of the peak that 
is strongly temperature dependent in the neighborhood of $T_{c}$. 
The intensity vanishes slightly above $T_{c}$, without any 
apparent change in the shape of the peak itself.  Indeed, the sharp 
temperature dependence of this intensity in the neighborhood of 
$T_{c}$ is consistent with its being proportional to a fractional 
power of the (local) superfluid density.  (See Eq. (\ref{eq:Zcrit}).)  
Additional evidence for this comes from an old observation of 
Harris {\it et al.}\cite{shen3} that, as a function of underdoping, the weight 
in the peak at low temperatures decreases with 
decreasing superfluid density. Moreover, Shen and Balatsky\cite{shenandb} have
argued that a small dispersion of the ARPES peak in the direction perpendicular
to the putative stripe direction scales more or less with $T_c$, consistent 
with our Eq. (\ref{eq:teff}).  The distance between the coherent peak
and the dip feature in ARPES curves near the ${\bar M}$ 
point \cite{campuzano} decreases with underdoping, consistent with 
the zero temperature distance being proportional to $T_c$.

A similar {\it temperature} evolution has been observed for the so called
``resonant peak'' in neutron scattering\cite{neutron} in {\YBCO} and {\BSCO}
(although no such feature has been seen\cite{notseen} in {\LSCO}).   
We would like to identify this phenomenon, as well, 
with a dimensional crossover of the sort discussed here.  
However, the spin resonance we have found in the present model is clearly not 
directly related to the observed resonant peak.  
In particular, all features we have found are peaked at a momentum $2k_F$, 
while the resonant peak is centered on the antiferromagnetic 
wave-vector $(\pi,\pi)$. Moreover, the peak we have found disappears through 
a lifetime catastrophe well below $T_c$, while the resonant peak is sharp 
immediately below $T_c$. Clearly, at the very least, to
have a theory with anything more than a very rough caricature of this observed
magnetic behavior, we need to expand the considered model\cite{cc} to include 
the effects of the antiferromagnetic ``strips'' between the ``stripes.''

\subsection{Further Implications for High Temperature Superconductors}

Finally, we end with a few additional observations concerning insights into 
the behavior of the high temperature superconductors that can be obtained 
from the analysis of this paper:
 
In the superconducting state of a structurally quasi-1D superconductor, 
we have found there are two emergent length scales, $\xi_s$ and $\xi_c$.  
If the quasi-1D electronic structure is self organized, as it is in the 
high temperature superconductors, there
are potentially two additional emergent length scales:\cite{zohar}  
the mean spacing between stripes,
$d$, and the persistence length of the stripes, $\xi_{stripe}$.  
$d$ can be determined
directly from  the charge incommensurability\cite{tranquada,mook} (or 
indirectly\cite{tranquada,oron} from the spin) structure factor. 
$\xi_{stripe}$ is much harder to determine experimentally, 
although it is bounded below by
the correlation length of the magnetic order.\cite{tranquada2,zachar2}  
So long as $\xi_{stripe}$
is the longest length scale in the problem, {\it i.e.} so long as
$\xi_{stripe}\gg\xi_c$, it is
possible to assume, as we have here, that the superconducting 
properties of the system are
quasi one-dimensional.  
Where this inequality is violated, the correct theory of the
superconducting state needs to be significantly modified.  
As was pointed out previously,\cite{spingap} so long as the weaker 
condition, $\xi_{stripe}\gg \xi_s$, is
satisfied, it is possible to have a one dimensional theory of 
spin gap formation.  At present
experiments are unclear about the range of doping and materials
for which either of these inequalities is satisfied, 
which is the most important
source of uncertainty in the application of these ideas to the 
high temperature superconductors.  
Certainly, with sufficient overdoping, the stripes loose their 
integrity and the application of these ideas becomes suspect.

To get a feeling for magnitudes, 
we can make rough quantitative estimates of the remaining
length scales from well-established experimental data in the high temperature
superconductors, although numbers vary from material to material, 
and as a function of doping
concentration, $x$.  
The spin velocity in the undoped antiferromagnet is around $v_s\approx
0.8$eV-\AA, and the superconducting gap is $\Delta_0\approx 35$meV, 
so $\xi_s\sim$ 20\AA. The charge coherence length
$\xi_c=\xi_s(v_c/v_s)(\Delta_s/\Delta_c)$, so if we  estimate
$v_c/v_s\sim t/J\approx 2-3$ and $\Delta_c\approx 2T_c\approx 16$meV, 
we find that characteristically $\xi_c\sim 100-150$\AA.  
(This is in good agreement with the $\mu$SR
measurement\cite{sonier} of the vortex core radius cited above.  )
The spacing between stripes
is in the range of four or more lattice constants, $d \sim 16$\AA.

A crossover magnetic field, which can be identified as a
mean field\cite{hc2}
$B_{c2}$, can be estimated as the field at which there 
is one vortex per coherence length
$\xi_c$ between each pair of neighboring stripes;  this leads to an estimate
\be
B_{c2} \sim \phi_0/\xi_c d \; ,
\ee
where $\phi_0=hc/2e$ is the superconducting flux quantum. 
While $B_{c2}$ estimated in this
fashion is quite large 
($\phi_0/\xi_c d=80T$ for $d = 16$\AA and $\xi_c=100$\AA) it is
small compared to the characteristic magnetic field, 
\be
B_s\sim 2\phi_0/\xi_s w \; ,
\ee
at which orbital effects lead to the destruction of the spin gap.  
Here, $w$ is the ``width''
of a stripe - {\it i.e.} 
the width of the 1d region involved in spin gap formation.  In the
``spin gap proximity effect'' mechanism\cite{spingap} proposed 
previously this would imply
that $w$ is one to two times the crystalline lattice constant.  
The extremely large value of
$B_s$ rationalizes the lack of any observable reduction of the spin gap 
temperature in the recent NMR experiments of 
Gorny {\it et al.}\cite{gorny} up to fields as high as 12T in
\YBCO.  (See, also, the discussion in Ref. \ref{ref:zachar}.)

One important difference between a stripe phase and the array of chains studied
here, is that in the stripe phase there are additional electronic degrees of
freedom which live in the antiferromagnetic strips between the stripes.  The
two-component nature of the electronic structure of doped
antiferromagnets,\cite{losalamos} is characteristic of the micro-phase
separation physics that gives rise to this state.  Of course, the
antiferromagnetic strips are themselves quasi one-dimensional
magnets,\cite{birgtheory} so that any magnetic ordering must be viewed, in
similar spirit to that considered here, as resulting from a dimensional
crossover.  Indeed, it is certainly the spins in the insulating strips that
make the dominant contribution\cite{cc} to the 
``resonant peak'' observed in neutron
scattering.  A detailed theory of this peak is beyond the scope of the 
present model, but is embodied in the spin gap proximity effect.%
\cite{spingap}

However, we have found a new neutron scattering resonance for a 
quasi one-dimensional superconductor.  While the dimensional crossover
causes no bound state in the spin-1 excitations, we find a resonant
state of two spin-${1 \over 2}$ quasiparticles appearing below 
$T \approx 0.4 T_c$.  The mode appears at an energy $2 \Delta_s + \Delta_c
= 2 \Delta_0 $, or twice the single particle gap as measured
by ARPES or tunneling, and at momentum $2 k_F$, where $k_F$ defines
the Fermi surface associated with a stripe.  Since this is 
a four soliton resonance, it may be qualitatively sensitive to 
deviations from the limit $\Delta_c << \Delta_s$, so that the
resonance is likely to be most well defined in the underdoped
region where $T_c\ll\Delta_0$.  

Finally, we remark that the ARPES spectrum along the symmetry 
direction from (0,0) to $(\pi,\pi)$, {\it i.e.} along the ray which 
is expected to pass through the node of a d-wave
gap function, is very different in character from that in the 
${\bar M}$ that we have discussed.  In clean samples of optimally 
doped \BSCO, there is a peak\cite{peak,valla} in the
spectral function both above and below $T_c$, 
and the peak reaches the Fermi surface at a
well defined ``nodal point'', $\vec k_n=(0.44\pi,0.44\pi)$.  
This peak does not exhibit the
characteristics of a quasiparticle peak,\cite{valla} in that its 
width is always larger than its energy;  indeed, it seems to exhibit 
quantum critical behavior reminiscent of a Luttinger Liquid.  
Moreover, there is no qualitative change\cite{valla} in the temperature
evolution of this peak as the temperature is lowered from two 
or three times $T_c$ down to
temperatures as low as at least 1/2 $T_c$;  
the character of the nodal excitation seems to be
remarkably insensitive to the onset of superconductivity.  
By contrast, in optimally doped
\LSCO there is apparently\cite{ino} no observable peak along 
the nodal direction, and indeed
little or no spectral weight within about 0.5 eV of the Fermi energy.  
Indeed, recent neutron
scattering studies\cite{aeppli} of the low energy magnetic scattering in the
neighborhood of
$2\vec k_n$ have revealed the existence of a clean spin gap 
at low temperatures, which is apparently inconsistent with 
the existence of any gapless nodal quasiparticle excitations.

It is clear that whatever spectral response is observed near $\vec k_n$ is 
not associated with the vertical and horizontal stripes studied here, because
a stripe wave vector does not span the ``Fermi surface'' 
along this direction.  
It could be associated with diagonal stripes, which have been observed 
recently in various insulating materials,\cite{diag1,diag2} in which case the 
observed quantum critical behavior might truly be that of a Luttinger liquid.  
An alternative picture is backflow associated with holes that have not
condensed into vertical or horizontal stripes. Both explanations are
conceivable as there are strong reasons to expect that the orienting 
potential, which locks the stripes along a particular (vertical or horizontal) 
crystallographic direction to be stronger in {\LSCO} than in {\BSCO}. 
However, other sources of quantum critical behavior are 
certainly possible.\cite{varma}  
We will defer further discussion of these classes of excitations to a future 
study.

\acknowledgements 
We would like to acknowledge useful discussions with E.~Fradkin, L.~Pryadko,
J.~Tranquada, Z.~X.~Shen, P.~B.~Wiegmann, A.~Tsvelik, O.~Zachar, 
and C.~Nayak.  
This work was supported in part by NSF grants number DMR98-08685 
(EWC and SAK), DMR-9814289 (SAK) and DOE grant number 
DE-AC02-98CH10886 (VJE). D.~O. has been supported by the Rothschild 
Fellowship. SAK gratefully acknowledges the
hospitality of the Physics and Applied Physics Departments at
Stanford University where much of this work was carried out.

\appendix

\section{The effect of interchain single particle hopping}
\label{intertun}

Until now, we have ignored single particle hopping between chains.  
This is because,
especially in the presence of a spin gap, it is irrelevant in 
the renormalization group sense.  However, in the superconducting state, 
we expect the quasiparticles to be able to
propagate coherently between chains.  
Because these terms are irrelevant, their effects on
the spectrum can be computed in ordinary degenerate perturbation theory.  
It is easy to see that to first order in the interchain hopping, 
the quasiparticle energy is
\be
\label{1stpert}
{\cal E}({\vec k})=\sqrt{v_s^2 k_\parallel^2+\Delta_0^2}+
Z(k_\parallel)\epsilon^{(\perp)}(\vec k_{\perp}) + 
{\cal O}\big( \epsilon^{(\perp)}\big)^2 \; ,
\ee
where $\epsilon^{(\perp)}(\vec k_{\perp})=2t_\perp\cos(k_\perp a)$ 
is the interchain contribution to the quasiparticle dispersion, 
and $k_\parallel+k_F$ and $\vec k_{\perp}$ are,
respectively, the components of the crystal momentum parallel 
and perpendicular to the chain direction.   

This is highly reminiscent of the spectrum we would have obtained were we
to compute the spectrum of a quasi one-dimensional superconductor 
using BCS mean field theory
\ba
{\cal E}_{\vec k}^{(BCS)}=&& \sqrt{[v_Fk+\epsilon^{(\perp)}(\vec
k_{\perp})]^2+\Delta^2} \\
=&&
{\cal E}_k^{(BCS)}+[v_sk/{\cal E}_k^{(BCS)}]\epsilon^{(\perp)}(\vec k_{\perp}) 
+{\cal O}\big(\epsilon^{(\perp)}\big)^2 \; ,
\nonumber
\ea
with the differences that the Fermi velocity is replaced by the 
(slower) spin velocity,
the superconducting gap is the sum of the (single-chain) spin 
gap and the (interchain)
charge gap, and the interchain bandwidth is reduced by the 
quasiparticle weight factor $Z$. 

\section{Macroscopic Superfluid Density}

In this appendix, we compute the macroscopic phase stiffness 
(superfluid density) tensor 
$K_{ab}[\kappa]$
in two dimensions $(a=x,y)$ 
given a microscopic distribution of the (in general anisotropic) 
local phase stiffness tensor,
$\kappa_{ab}(\vec r)$.  
We include the derivation here for pedagogical purposes, although 
the results exist elsewhere in the literature \cite{Bergman92}.

$\kappa$ determines the relation between the local current density, 
$\vec j(\vec r)$ and the gradient of the phase according to
\be
j_a(\vec r)=\kappa_{ab}\partial_b\theta(\vec r) \; .
\ee
From the equation of continuity, it follows that 
$\vec\nabla \cdot \vec j=0$, so we can express
$\vec j$ in terms of a potential, $j_a(\vec r)=
\epsilon_{ab}\partial_b\phi(\vec r)$, so that
\be
\epsilon_{ab}\partial_b\phi(\vec r)= \kappa_{ab}\partial_b\theta(\vec r) \; .
\ee
To compute $K_{xx}$ in a rectangular geometry, 
this equation is to be solved subject to the
boundary conditions that $\theta=0$ for $x=0$ and 
$\theta=\Delta\theta$ for $x=L_x$ (independent
of $y$) and (from the condition that no current 
can flow out of the sample in the $y$
direction) $\phi=0$ for $y=0$ and $\phi=\Delta\Phi$ for $y=L_y$.  
For a given distribution of $\kappa$, we solve this equation for 
given $\Delta\theta$ to determine $\Delta\phi$, from which
we determine $K$ according to
\be
K_{xx}[\kappa]=\Delta\phi/\Delta \theta \; .
\label{eq:Kxx}
\ee 

The key observation is the same potential and phase that satisfy Eq.
(\ref{eq:Kxx}), also satisfy the dual equation
\be
\epsilon_{ab}\partial_b\theta(\vec r)= \kappa^D_{ab}\partial_b\phi(\vec r) \; ,
\ee
where
\be
\kappa^{D}_{ab}(\vec r)\equiv\epsilon_{ac}\kappa^{-1}_{cd}(\vec r)
\epsilon_{db} \; .
\ee
Therefore
\be
K_{xx}[\kappa]K_{yy}[\kappa^D]=1 \; .
\ee

We can apply this general result to the problem of interest here.  
Consider the case of a
square geometry in which, 
because of some assumed domain structure, the system is macroscopically
isotropic ($\bar \kappa\equiv K_{xx}=K_{yy}$) despite the existence of 
microscopic anisotropy in each "stripe" domain.  It follows that
\be
\bar\kappa(\kappa_{\perp},\kappa_{\parallel})\,
\bar\kappa(1/\kappa_{\parallel},1/\kappa_{\perp})=1 \; .
\label{eq:constr}
\ee
It follows that 
$\bar\kappa(\kappa_{\perp},\kappa_{\parallel})=\sqrt{
\kappa_{\perp}\kappa_{\parallel}}$.
Other solutions to  Eq. (\ref{eq:constr}) exist, 
but are not homogeneous functions.

\section{The Effective Potential, $v(x)$}

To compute the effective potential which appears in Eq. (\ref{eq:Heff}), we 
consider the discrete version of the refermionized Hamiltonian\cite{kllpy},
\be
H=-\sum_n[t_0+(-1)^n\Delta(n)/2][c_n^{\dagger}c_{n+1}+{\rm H.C.}] \; ,
\ee
where $v_c =2t_0$ and 
\be
\Delta(n)=\Delta_c {\rm sign}(R^2-4n^2) \; ,
\ee
corresponding to a pair of solitons separated by a distance $R$.  (We have
set the lattice constant equal to 1.)  

We compute the ground state energy on a system of $2N$ sites by computing the
single particle eigenvalues, and then summing over the lowest lying
$N-1$ of them to get the total energy as a function of $R$.  This is precisely
the program carried out previously to study various solitonic states in the
SSH model of polyacetylene.\cite{chx,ssh,tlm}  With open boundary conditions,
the Hamiltonian matrix is tridiagonal, and so particularly simple to study 
numerically for large system sizes.  We have carried out this 
program numerically for system sizes up to $2N=3000$, 
and for $\Delta_c=$ 0.2, 0.1, 0.05, and 0.02;  
the continuum limit is obtained when $\Delta_c\rightarrow 0$.  
Even for the smallest values of $\Delta_c$, we find no significant finite 
size effects at these large system sizes.  The results for $v(x)$ computed 
in this way are summarized in Fig. 1. The fact that the asymptotic value of 
$v$ is always slightly negative is a reflection of the fact that in the limit 
of small soliton width (which is equal to 1, the lattice constant, 
in the present calculation), the soliton creation energy is a very strongly 
varying function of the width, as found previously by Takayama, Lin-Liu, 
and Maki\cite{tlm}, and only approaches its true asymptotic limit 
$\Delta_c/2$, when $1/\xi_c$ is extremely small.

\end{multicols}


\begin{references}
\bibitem{ideology}  There is a direct sense in which the Luther-Emery 
model can be viewed as the fixed point of the massive sine-Gordon 
model.  While in any renormalization group scheme which is 
perturbatively related to a momentum shell RG transformation on the 
bosonic fields the flows pass through the Luther-Emery 
line, if we define a momentum-shell renormalization group 
transformation in the refermionized form, with field rescaling chosen 
to preserve the mass term, then the four-fermion interaction is 
manifestly irrelevant, and indeed even the kinetic energy is a 
(dangerously) irrelevant interaction.
\bibitem{meanfield}  The interchain mean field approximation 
will always introduce errors associated
with the presence of superconducting phase 
modes with long wavelength in the direction transverse to
the chain direction.  They may change the analytic form of 
the threshold behavior of the
various incoherent response functions, but as they are generally weakly 
coupled, they generally\cite{notagap} do not have any
strong quantitative effect.
\bibitem{astute} Note that there is a trouble with 
the equation of continuity implied by these statements.  
This is the same problem that occurs in any mean field treatment of a
superconducting state, which derives from the fact that the mean 
field breaks local gauge invariance.  In a more complete treatment, 
a backflow of supercurrent on neighboring chains ensures the validity of 
the equation of continuity, but since this
flow is associated with a current of cooper pairs, 
it leaves the parity of $N_c$ a sharply defined quantity.\cite{rokhsarandme}
\bibitem{notagap}  Although in the context of the interchain 
mean field theory employed here, $\Delta_{c}$ appears as a true gap 
for charged excitations, in a more complete treatment it is simply a 
characteristic energy scale.  Fluctuations of the superconducting 
phase with long wavelengths in both the transverse and longitudinal 
directions, the Goldstone modes of the superconducting order, have 
energies which, at least in the absence of long range Coulomb 
interactions, extend all the way down to zero energy.  $\Delta_{c}$ is 
actually something like a Debye energy for excitations of the 
superconducting phase with short wavelength in the interchain 
directions.  Similarly, below $T_{c}$, $\xi_{c}$ is actually a 
Josephson length.
\end{references}
\end{document}